\documentclass[twocolumn,pra,showpacs,amsmath,amssymb]{revtex4}
\usepackage{graphicx}
\usepackage{bm}
\usepackage{epstopdf}
\usepackage{epsfig}
\usepackage{dsfont}
\usepackage{amsmath,amsfonts,amssymb}
\usepackage{graphicx}
\usepackage{physics}
\usepackage{mathtools}
\usepackage{cancel}
\usepackage{wrapfig}
\usepackage[utf8]{inputenc}
\usepackage{epigraph}
\usepackage{tcolorbox}
\usepackage{titlesec}
\usepackage{dsfont}

\begin{document}

\title{Simplest non-additive measures of quantum resources}
\author{L. F. Melo and Fernando Parisio}
\email[]{parisio@df.ufpe.br}
\affiliation{Departamento de
F\'{\i}sica, Universidade Federal de Pernambuco, Recife, Pernambuco
50670-901 Brazil}

\begin{abstract}
Given an arbitrary state $\rho$ and some figure of merit ${\cal E}(\rho)$, it is usually a hard problem to determine the value of ${\cal E}(\rho^{\otimes N})$.
One noticeable exception is the case of additive measures, for which we 
simply have ${\cal E}(\rho^{\otimes N}) = Ne$, with $e\equiv {\cal E}(\rho)$. In this work we study measures that can be described
by ${\cal E}(\rho^{\otimes N}) =E(e;N) \ne Ne$, that is, measures for which the amount of resources of $N$ copies is still determined by the single real variable $e$, 
but in a nonlinear way. If, in addition, the measures are analytic around $e=0$, recurrence relations can be found for the Maclaurin 
coefficients of $E$ for larger $N$. As an example, we show that the $\ell_1$-norm of coherence is a nontrivial case of such a behavior.
\end{abstract}
\maketitle

\section{Introduction}
\label{intro}
Several tasks of practical utility in quantum information science require access to a number of copies of 
conveniently prepared states $\rho$. If the ability to succeed in a particular task, given that we posses a single copy of $\rho$, can be quantified 
by some function ${\cal E}(\rho)$, it turns out that only very seldom ${\cal E}(\rho^{\otimes N})$, the same ability for $N$ copies, depends solely on ${\cal E}(\rho)$ in a linear way. 
That is ${\cal E}(\rho^{\otimes N}) \ne N{\cal E}(\rho)$, in general. 
This fact poses a difficulty in the determination of ${\cal E}(\rho^{\otimes N})$ because its direct evaluation requires computations, 
and often optimizations, in Hilbert spaces whose dimensions grow exponentially as a function of $N$.

This difficulty is one of the main motivations for the pursuit of asymptotic results, which are useful 
whenever one can assume that the number of available states is so large that taking the limit
$N\rightarrow \infty$ is a justifiable approximation.
However, it is not unusual that, even for regularizable measures, the asymptotic regime dominates only for an
impracticably high number of copies \cite{oneshot,natcomm,ieee}. 
For this reason, one cannot always evade the problem
of evaluating quantum figures of merit related to a large, but finite number of copies. 

Of course, there is no such a difficulty when the considered quantifier is additive, as, for instance, the squashed entanglement \cite{squashed}  and the logarithmic negativity \cite{vidal} (both non-separability quantifiers). Additivity, however, is a very restrictive condition. In a recent work \cite{parisio} one of us characterized possible measures which can be considered as generalizations
of additive measures,  in two different ways. These ``scalable'' measures are such that ${\cal E}(\rho^{\otimes N})=E^{(N)}(e_{i_{1}}, e_{i_{2}}, \dots, e_{i_{q}})$,
with $e_j\equiv {\cal E}(\rho^{\otimes j})$, $j<N$. That is, the amount of the resource \cite{comm0} for $N$ copies is completely determined by the $q$ real numbers given
by the amount of the same resource for a set of some smaller numbers of copies. We call quantifiers satisfying this property $q$-Scalable ($q$-S). 
While the additive relation ${\cal E}(\rho^{\otimes N}) = Ne$ is a (i) {\it linear} function
of a (ii) {\it single} real variable, a general $q$-scalable measure is a {\it nonlinear} function of {\it several} ($q$) real variables. In \cite{parisio} constraints for a function to be a valid scalable measure were presented and recurrence relations for the Maclaurin series of all physically consistent 1-S functions ${\cal E}(\rho^{\otimes N})$ were determined.

In the present work we focus on the possible nonlinearity of quantum measures that, however, still depend on the single real variable $e$ (1-S measures). These are, arguably, the simplest 
behaviors displayed by measures of quantum resources. To exemplify our results we show that the $\ell_1$-norm of coherence \cite{coh} is a non-trivial 1-scalable measure, presenting a distinct form of scalability when compared with other coherence quantifiers in the literature.
\section{1-Scalable measures}
We will denote measures which are functions of $e$ and $N$ only, through 
\begin{equation}
\label{def0}
\mathcal{E}(\rho^{\otimes N})=E^{(N)}\left(\mathcal{E}(\rho)\right)=E^{(N)}(e).
\end{equation}

By definition, the condition $E^{(1)}(e)=e$ must be satisfied. For any quantifier we will assume that $\mathcal{E}(\rho^{\otimes N})=0$ 
for zero-resource states $\rho$ (but not the other way around). Note that, while ${\cal E}: {\cal B(H)}^{\otimes N} \mapsto \mathds{R}_{+}$, we have
$E:\mathds{R}_{+}\mapsto  \mathds{R}_{+}$, so that, typically, 
{\it the domain of the latter has a dimension which is much lower than that of the former.}

The following auxiliary definitions will be useful. Given an arbitrary positive integer $a$, one can take $N=a^n$ and $K=a^k$, with
$n,k \in \{0,1,2,\dots\}=\mathds{N}$, $k\le n$, such that $N,\,K \in \mathds{P}_{a}$,
\begin{equation}
\mathds{P}_{a}=\{1,a, a^2, \dots\}.
\label{set}
\end{equation}
It has been shown that 1-S measures must satisfy:
\begin{equation}
\label{core}
E^{(N)}(e)=E^{(N/K)}\left(E^{(K)}(e)\right).
\end{equation}   
This constraint is sufficient to determine all possible forms of analytic 1-S measures and comes 
from the requirement that the tensor product structure must be preserved by any consistent figure of merit \cite{parisio}.
\section{Analytic 1-S measures}
In addition to the consistency relation (\ref{core}), we will assume that $E^{(N)}(e)$ is analytic at $e=0$. More explicitly, consider that the function
$E^{(N)}(e)$ has a Maclaurin series that converges in the non-vanishing interval $[0, \epsilon_N)$, $\epsilon_N>0$, 
$E^{(N)}(e)=\sum_{j=1}^{\infty}d_j(N)\, e^j$,
for $e \in [0, \epsilon_N]$. We defined $d_j(N)=\frac{1}{j!}\frac{d^jE^{(N)}}{de^j}\arrowvert_{e=0}, $
with  $d_0(N)=E^{(N)}(0)=0$ because $e=0 \Rightarrow {\cal E}=0$ and $d_j(1)=\delta_{1,j}$ because $E^{(1)}(e)=e$.
In \cite{parisio}, it has been demonstrated that once one knows the Maclaurin coefficients $d_j(K)$ for some $K \in \mathds{P}_{a}$, then, the coefficients 
for the series associated with a larger number of copies $N$ can be found. The general recursive relation satisfied by the Maclaurin coefficients $d_j(N)$ of 1-S analytic (at $e=0$) 
measures is given by
\begin{equation}
\label{coef}
d_j(N)=\sum_{\ell=1}^j d_{\ell}(N/K)\sum_{i=1}^{{j-1}\choose{\ell-1}}\pi_{i}(j,\ell;K),
\end{equation}
where $\pi_{i}(j,\ell;K)=d_{\mu^i_1}(K)\,d_{\mu^i_2}(K)\dots d_{\mu^i_{\ell}}(K)$
is a product with $(\mu^i_1, \mu^i_2, \dots, \mu^i_{\ell})$ being the $i$-th composition of $j$ into $\ell$ parts.
A composition of an integer $j$ in $\ell$ parts is an {\it ordered} sum $j=\mu_1+\mu_2+\dots+\mu_{\ell}$,
of strictly positive integers. 
A well-known result in enumerative combinatorics is that there are ${j-1}\choose{\ell-1}$ such compositions \cite{comb}.
To be precise, in \cite{parisio}, the general recurrence relation has been demonstrated for infinite Maclaurin series. Here,
we note that this is not the most general situation under which (\ref{coef}) is valid.

If we suppose that the series is finite, with $E^{(N)}(e)$ being a polynomial of arbitrary degree, $E^{(N)}(e)=\sum_{k=1}^{L(N)} d_k(N)e^k$, we can show that (\ref{coef}) remains valid, provided that the upper limit $L(N)$ satisfies the constraint derived in what follows. Using property (\ref{core}) we get:
\begin{eqnarray}
\sum_{k=1}^{L(N)} d_k(N)e^k&=&\sum_{l=1}^{L(N/K)}d_l(N/K) \left[\sum_{m=1}^{L(K)}d_m(K)e^m\right]^l. \label{firstmaclaurin}
\end{eqnarray}
The number of terms in each side of this equality corresponds to the largest power of $e$, which, in the left-hand side is $L(N)$, while it is $L(N/K)L(K)$, in the right-hand side. Therefore, whenever $E^{(N)}(e)$ is a polynomial of arbitrary, but finite degree, we must have:
\begin{eqnarray}
L(N) = L(N/K)L(K). \label{L(N)}
\end{eqnarray}
Choosing $K=a$ (remember that $N \in \mathds{P}_{a}$) and changing the notation $L(N)=L(a^n)$ to $L_n$ we get the recurrence $L_n=L_{n-1}L_1$, which has a simple solution:
\begin{equation}
L(N)=[L(a)]^{\log_{a}N}= [L(a)]^n, \label{L solution}
\end{equation}
where $L(a)$ is the upper limit of the series expansion of $E^{(a)}(e)$. If $L(a)\rightarrow \infty$ then, of course, $L(N)\rightarrow \infty$.

Note that each recurrence $d_j(N)$ depends on the previous relations $d_{l<j}(N)$. The general form of the first and second order coefficients have been derived in \cite{parisio}. As an illustration of how (\ref{coef}) works, we obtain the next, 3rd order coefficient. In this case, the recurrence relation reads:
\begin{eqnarray}
d_3(N) &=& d_1(N/K)\pi_1(3,1,K)+ d_3(N/K)\pi_1(3,3,K)  \nonumber\\ 
&+& d_2(N/K) \sum_{i=1}^{2} \pi_i(3,2,K). \nonumber
\end{eqnarray}
Using the compositions of 3 into 1 and 3 terms, we have $\pi_1(3,1,K)=d_3(K)$ and $\pi_1(3,3,K)=[d_1(K)]^3$. As we can only sum 3 via the pairs 1+2 and 2+1, $\pi_1(3,2,K)=\pi_2(3,2,K)=d_1(K)d_2(K)$, then we get:
\begin{eqnarray}
d_3(N) &=& d_1(N/K)d_3(K) + d_3(N/K)[d_1(K)]^3\nonumber\\
&+& 2d_2(N/K)d_1(K)d_2(K). \label{3r}
\end{eqnarray}
The solution of this recurrence relation is given in appendix \ref{appendix:3order}, and the result along with the lower order solutions, leads to
\begin{eqnarray}
E^{(N)}(e) &=& N^\nu e + d_2(a)\left(\frac{N}{a}\right)^\nu \left(\frac{N^\nu-1}{a^\nu-1} \right)e^2 \nonumber \\
&+& \left \lbrace 2[d_2(a)]^2 \left(\frac{N}{a^2}\right)^\nu \frac{(N^\nu-1)(N^\nu-a^\nu)}{(a^{\nu}-1)(a^{2\nu}-1)}\right. \nonumber\\
&+& \left.d_3(a)\left(\frac{N}{a}\right)^{\nu}\left(\frac{N^{2\nu}-1}{a^{2\nu}-1}\right) \right \rbrace e^3+ O\left(e^4\right),
\label{maclaurin}
\end{eqnarray}
where $\nu=\log_a(d_1(a))$. Note the meaning of this result: Any general figure of merit $\mathcal{E}(\rho^{\otimes N})$ which can be written as an analytic function of $\mathcal{E}(\rho)=e$ and $N$ must satisfy the above relation.

In principle, the iteration can be continued up to arbitrary order, possibly with the aid of numeric computations, in the general case. We recall that the important conceptual result is that if we know the expansion for $a$ copies $E^{(a)}(e)=d_1(a)e+d_2(a)e^2+\dots$, the coefficients in (\ref{maclaurin}) are determined. If one knows how the measure works for, say, 3 copies of a system there is a systematic way to determine the behavior of this measure for $N=3^n$ copies.

\begin{figure}
\resizebox{0.45\textwidth}{!}{
 \includegraphics{{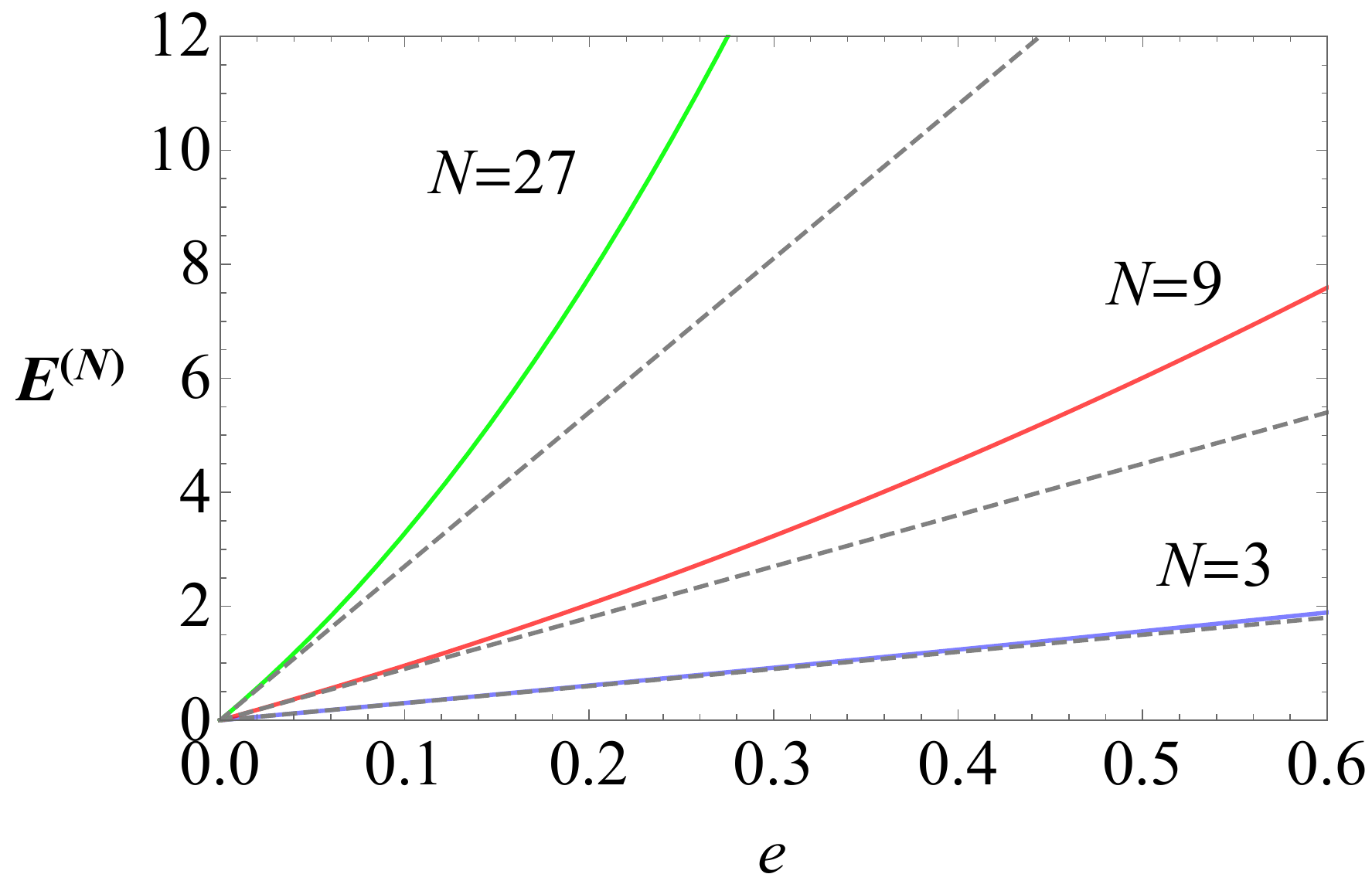}}
}

\caption{Plots of the functions $E^{(3)}(e)=3\,(e+\delta  e^2+\delta^2  e^3)+ \dots \approx 3\,e$, $E^{(9)}(e)$, and $E^{(27)}(e)$, for $\delta=0.08$. The small nonlinearity displayed by $E^{(3)}(e)$ causes a much more pronounced deviation from linearity in  $E^{(9)}(e)$ and $E^{(27)}(e)$.}
\label{fig1}
\end{figure}

From (\ref{maclaurin}), it is clear that any nonlinearity in $E^{(a)}(e)$ is enhanced for a larger number of copies. As an illustration of this behavior, consider the hypothetical, slightly nonlinear dependence $$E^{(a)}(e)=a\,(e+\delta e^2+\delta^2  e^3)+ \dots \approx a\,e,$$ with $\delta<<1$, that is, $d_j(a)=\delta^{j-1} a$. By inserting these coefficients into (\ref{maclaurin}) we get the functional dependencies for larger $N$s, up to third order. In Fig. (\ref{fig1}) we set $a=3$ and plot the functions $E^{(3)}(e)$, $E^{(9)}(e)$, and $E^{(27)}(e)$, for $\delta=0.08$.
We see from (\ref{maclaurin}) that the condition for subadditivity (superadditivity) is $\nu<1$ ($\nu>1$). Note that for $\nu=1$ and $d_{j}(N)=\delta_{j1}$ we recover the additive case. In the next section we study one particular kind of 1-S measure.
\subsection{Two-coefficient case} 
\label{subsection:1,2}

If a quantum function $\mathcal{E}(\rho^{\otimes a})$ is 1-S and has the form $E^{(a)}(e)=d_1(a)e+d_2(a)e^2$ (only the first two coefficients are non-zero), then the series for $N$ copies ends at $L(N)=2^n$ [see equations (\ref{set}) and (\ref{L solution})]. With this simplifying hypothesis, the next higher order recurrence relations becomes:
\begin{eqnarray}
d_4(N) &=& d_2(N/a)[d_2(a)]^2 + 3d_3(N/a)[d_1(a)]^2d_2(a)  \nonumber  \\ 
&+& d_4(N/a)[d_1(a)]^4.\nonumber
\end{eqnarray}

For this particular case, we obtained the 4th order coefficient in closed form (we handled it with the help of the software \textit{Mathematica}):
\begin{eqnarray}
 \label{2coefficients}
d_4(N) &=& [d_2(a)]^3 \left(\frac{N}{a^3}\right)^{\nu} \left(\frac{N^{\nu} -1}{(a^{\nu}-1)^2}\right) \nonumber\\
&\times& \left(\frac{N^{\nu}-a^{\nu}}{a^{2\nu}-1}\right) \left(\frac{N^{\nu}(5+a^{\nu})-1-5a^{2\nu}}{1+a^{\nu}+a^{2\nu}}\right).
\end{eqnarray}
Unfortunately, there is no clear pattern suggesting how $d_k(N)$ should look like. However, if in addition to  $E^{(a)}(e)=d_1(a)e+d_2(a)e^2$, we assume $\nu=1$ and $a=2$, one gets the following series:
\begin{eqnarray}
E^{(N)}(e) &=& Ne + [d_2(2)]\frac{N}{2}(N-1)e^2 \nonumber \\
&+& [d_2(2)]^2 \frac{N(N-1)(N-2)}{1.2.3}e^3 \nonumber \\ 
&+& [d_2(2)]^3 \frac{N(N-1)(N-2)(N-3)}{1.2.3.4}e^4 \nonumber\\
 &+& O(e^5),
 \label{induced}
\end{eqnarray} 
ending at $L(N)=2^{\log_2 N}=N$. Although this might look a too restrictive situation, we will see, in the next section, that this will be sufficient to disclose the nontrivial 1-S character of a well-known coherence quantifier. 
Observing this pattern we suspect that the series (\ref{induced}) is constructed in terms of Newton coefficients $d_k(N) = [d_2(a)]^{k-1}{N \choose k}$. This is indeed the case and we summarize the result as the following proposition.

\textbf{Proposition} Let $N \in\mathds{P}_{2}$, $\rho \in \mathcal{B}(\mathcal{H})$ and $E$ be an analytic 1-S measure with respect to $\rho$ and $\mathds{P}_2$. If $\mathcal{E}(\rho^{\otimes 2})=E^{(2)}(e)$ satisfies $E^{(2)}(e)=2e+[d_2(2)]e^2$, exactly, then:
\begin{equation}
E^{(N)}(e) = \sum_{k=1}^{N}{N \choose k} [d_2(2)]^{k-1} e^{k} \label{binomial}
\end{equation}
with $e=\mathcal{E}(\rho)$ (the demonstration is given in appendix \ref{bicoeff}).

\section{Scalability of the $\ell_1$-norm coherence}
Coherence is a fundamental resource in a variety of tasks and, often enhanced with the use of quantum systems. Naturally, many ways to quantify it
have been proposed in the last decades \cite{coh, coh2, coh3, coh4, coh5, coh6, coh7}.
In this section we give a non-trivial example of a coherence quantifier which is described by a 1-S function,
namely, the $\ell_1$-norm of an arbitrary state $\rho$, relative to a particular basis \cite{coh}. Given the matrix representation of the 
state in the chosen basis, the $\ell_1$-norm corresponds to a sum involving all off-diagonal entries:
\begin{equation}
{\cal C}_{\ell_1}(\rho)\equiv\sum_{i \ne j}^{d}|\rho_{ij}|=\sum_{ij}^{d}|\rho_{ij}|-1\equiv c. \label{defc}
\end{equation}
No assumption is made on the form of $\rho$ and on the dimension $d$, which we suppress in the remainder of this section. 
The sum after the second equality is over all possible values of the indexes, 
the equality being valid because ${\rm Tr}(\rho)=1$. This form
of expressing ${\cal C}_{\ell_1}$ will be particularly convenient for our purposes, since it immediately allows us to write
\begin{equation}
{\cal C}_{\ell_1}(\rho^{\otimes 2})=\sum_{ijkl}| \rho_{ij}\rho_{lk}| -1.\nonumber
\end{equation}

But $\sum_{ijkl} |\rho_{ij}\rho_{lk}|=\sum_{ij} |\rho_{ij}| \sum_{lk} |\rho_{lk}|=({\cal C}_{\ell_1}(\rho)+1)^2$. Therefore,
\begin{equation}
{\cal C}_{\ell_1}(\rho^{\otimes 2})=C^{(2)}_{\ell_1}(c)= (c+1)^2-1=2c+c^2.\nonumber
\end{equation}

This particular result has been obtained by Maziero \cite{maz}. Thus, $C^{(2)}_{\ell_1}(c)$ is exactly of the form of $E^{(2)}(e)$ in the proposition of the previous section, which leads to the hypothesis that $C_{\ell_1}$ is a 1-S function. Indeed, one can write the $\ell_1$-norm of coherence for $N$ copies as:
\begin{equation}
{\cal C}_{\ell_1}(\rho^{\otimes N})=\sum_{\begin{array}{c}i_{1},...,i_{N}\\j_{1},...,j_{N}\end{array}}|\rho_{{i_{1}}{j_{1}}}...\rho_{{i_{N}}{j_{N}}}|-1.
\end{equation}

\begin{figure}
\resizebox{0.45\textwidth}{!}{
 \includegraphics{{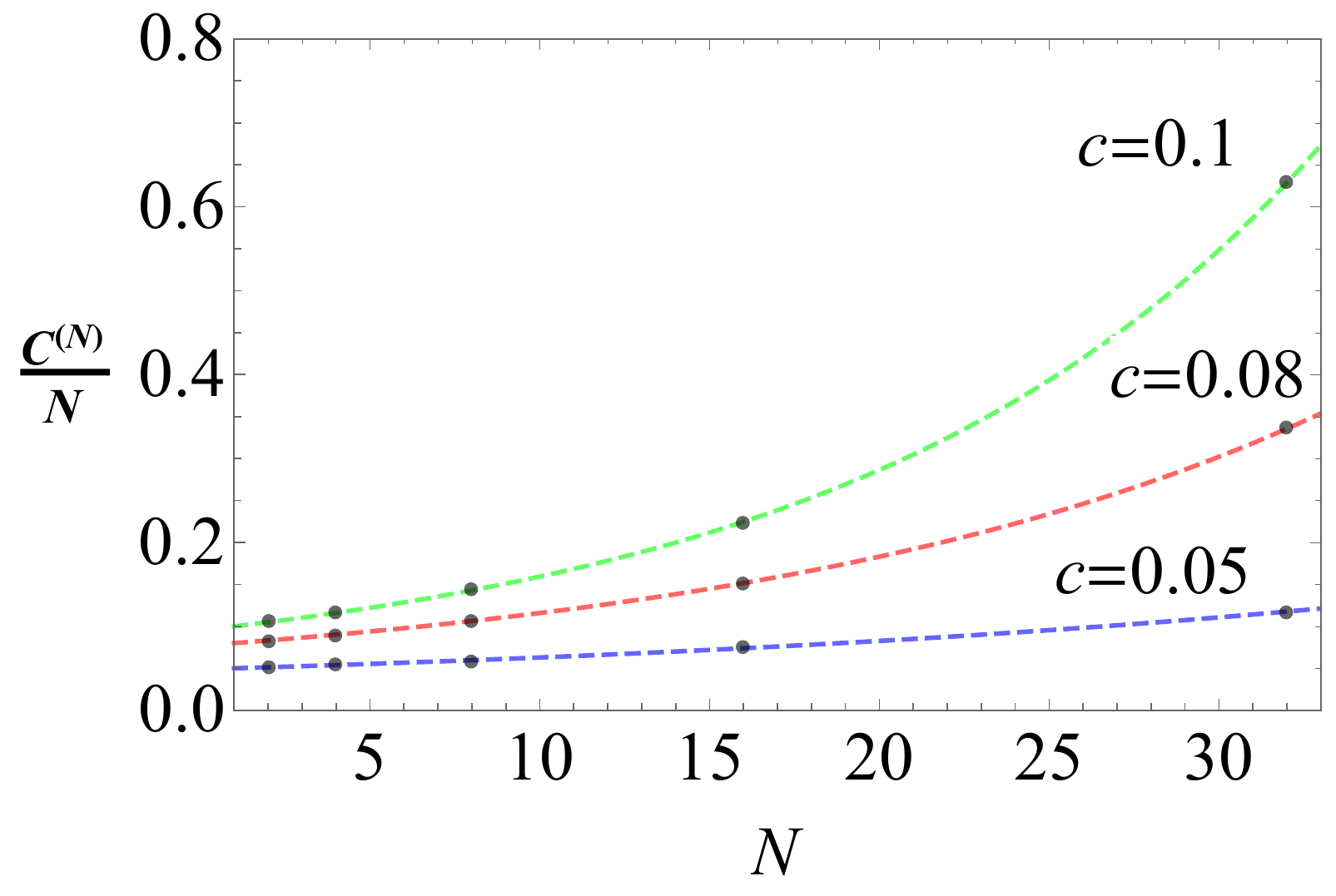}}
}
\caption{The $\ell_1$-norm of coherence per copy as a function of $N$. In this plot we have set the coherence of a single system to $c=0.05$ (lower bullets), $c=0.08$ (middle bullets), and $c=0.1$ (upper bullets). The dashed curves connect points corresponding to the same value of $c$.}
\label{fig2}
\end{figure}

Taking all the combinations of the indexes, the sum of the products is the product of the sums by taking the index $l$ varying from $0$ to $N$:
\begin{eqnarray}
{\cal C}_{\ell_1}(\rho^{\otimes N})&=&\prod_{l=0}^{N}\sum_{{i_{l}}{j_{l}}}^{d}|\rho_{{i_{l}}{j_{l}}}|-1 \nonumber
\end{eqnarray}
As all matrices are equal, this is simply equivalent to:
\begin{eqnarray}
{\cal C}_{\ell_1}(\rho^{\otimes N}) = {\left(\sum_{i,j}^{d}|\rho_{ij}|\right)}^{N} - 1,\nonumber
\end{eqnarray}
Therefore, for $N$ qudits we have 
\begin{equation} 
C^{(N)}_{\ell_1}(c)={\left(1+c\right)}^{N} - 1= \sum_ {k=1}^{N} 
{N \choose k}{c}^{k}. \label{scalabilityl1}
\end{equation}
We conclude that the $\ell_1$-norm of coherence of an arbitrary number of copies scales as a binomial series, and is, in fact, described by a 1-scalable function of $c$ [see fig.(\ref{fig2})]. Note that this measure is non-additive and diverges for a large $N$. This result is valid for qudits of arbitrary dimension and is in agreement with the proposition presented in the previous section for $d_2(2)=1$.

\subsection{The $\ell_2$-norm is not scalable} \label{chapter:l2}

Already in reference \cite{coh} (see its supplemental material) it was shown that the $\ell_2$-norm of $\rho$: 
\begin{equation}
\mathcal{C}_{\ell_2}(\rho) = \sum_{i \neq j}^d |\rho_{ij}|^2 \label{l2},
\end{equation}
could not possibly be a coherence measure because it violates the monotonicity condition under selective measurements, on average. Although this is, of course, sufficient to rule out $\mathcal{C}_{\ell_2}$ as a proper coherence measure, here we show, in addition, that this quantity is, in general, not scalable. Consider an arbitrary qubit state:
\begin{equation}
\rho \text{  } = \text{  }\left(\begin{matrix}
a & b \\
b^* & 1-a
\end{matrix} \right) \text{  }\text{ } \Rightarrow \text{  }\text{ } \mathcal{C}_{\ell_2}(\rho) = 2|b|^2 \equiv c_{2},\nonumber
\end{equation}
We use $c_{2}$ for the $\ell_2$-norm of $\rho$ to avoid confusion with (\ref{defc}). Using definition (\ref{l2}), we conclude that for, two copies:
\begin{eqnarray}
\mathcal{C}_{\ell_2}(\rho \otimes \rho) &=& 4\left[a^2|b|^2 + |b|^4 + |b|^2(1-a)^2 \right] \nonumber \\
&=& c_{2}^2 + 2[a^2 + (1-a)^2]c_{2}, \label{cl2}
\end{eqnarray}
For qubits, the sum of the squared modulus of all elements is equal to $a^2 + (1-a)^2 + 2|b|^2=\mathfrak{p}$, where ${\rm Tr} \rho^2\equiv \mathfrak{p}$ is the purity of $\rho$ (note that the relation $\mathfrak{p}-c_{2}\leq 1$ must be satisfied), then we can write (\ref{cl2}) as:
\begin{equation}
\mathcal{C}_{\ell_2} (\rho \otimes \rho) = (2 \mathfrak{p}) c_{2} - c_{2}^2 \ne \mathcal{C}_{\ell_2}^{(2)}(c_{2}) \label{l2norm2}.
\end{equation}
Differently from the $\ell_1$-norm, the $\ell_2$-norm for two copies depends not only on $c_{2}$, but also on $\rho$'s purity.
Due to this extra dependence, the $\ell_2$-norm of $\rho$ is, in general, not scalable. This analysis holds for any number of copies larger than 1.
To see this, let us split the sum $\sum_{l \neq k} |\rho_{lk}|^2$ for $\rho^{\otimes N}$ into two parts:
\begin{eqnarray}
\mathcal{C}_{\ell_2}(\rho^{\otimes N}) &=& \sum_{\begin{tiny}\begin{array}{c}i_1...i_N \\ j_1...j_N \end{array}\end{tiny}} |\rho_{i_1j_1}...\rho_{i_Nj_N}|^2 - \sum_{i_1...i_N}|\rho_{i_1i_1}...\rho_{i_Ni_N}|^2, \nonumber
\end{eqnarray}
and, similarly to the calculation in the previous section, we can rewrite the products as powers, and, thus:
\begin{equation}
\mathcal{C}_{\ell_2}(\rho^{\otimes N}) = \left( \sum_{ij}^d |\rho_{ij}|^2\right)^N - \left(\sum_{l}^d|\rho_{ll}|^2\right)^N.
\end{equation}
Using again the relation $a^2 + (1-a)^2 + 2|b|^2=\mathfrak{p}$ and recalling that the second sum is simply $a^2+(1-a)^2$, the difference between the trace of $\rho^2$ and $2|b|^2=c_{2}$, so:
\begin{equation}
\mathcal{C}_{\ell_2}(\rho^{\otimes N})=\mathfrak{p}^N - \left(\mathfrak{p} - c_{2}\right)^N, \nonumber
\end{equation}
where $c_{2}=\mathcal{C}_{\ell_2}(\rho)$. Even if we restrict the analysis to states of fixed purity, $\mathfrak{p}=\mathfrak{p}_0$, the scalability condition (\ref{core}) is not fulfilled (except for $\mathfrak{p}_0=1$):
\begin{eqnarray}
\mathcal{C}_{\ell_2}^{(N/K)}[\mathcal{C}_{\ell_2}^{(K)}(c_{2})] = \mathfrak{p}_0^{N/K} - \left(\mathfrak{p}_0-\mathcal{C}_{\ell_2}^{(K)}(c_{2})\right)^{N/K} \nonumber \\
= \mathfrak{p}_0^{N/K} - [\mathfrak{p}_0-(\mathfrak{p}_0^K-(1-c_{2})^K)]^{N/K}\nonumber \\
 \neq \mathcal{C}_{\ell_2}^{(N)}(c_{2}).\nonumber 
\end{eqnarray}
In the limited situation of pure states ($\mathfrak{p}_0=1$) $\mathcal{C}_{\ell_2}$ becomes sacalable, where we would have exactly the form (\ref{binomial}) with $d_2(2)=-1$ and the $\ell_2$-norm would be 1-S. In particular, note that equation (\ref{binomial}) indicates that the $\ell_2$-norm per copy for pure states would actually vanish for a large $N$ (in opposition to (\ref{scalabilityl1})):
\begin{equation}
\mathcal{C}^{(N)}_{\ell_2}(c_{2}) |_{\textbf{pure}} = 1 - (1-c_{2})^N = \sum_ {k=1}^{N} (-1)^k {N \choose k}{c_{2}}^{k}. \nonumber
\end{equation}
\section{Closing remarks}
In this work we characterized all possible quantum figures of merit $\mathcal{E}(\rho^{\otimes N})$ which can be expressed solely in terms of $\mathcal{E}(\rho)=e$ and $N$, as analytical functions of $e$. These are, arguably, the simplest measures after the additive ones. Although 1-S functions are far from encompassing all possibilities for resource quantification, for instance, 
1-S functions are unable to produce superactivation \cite{parisio}, they represent a much broader class of functions in comparison to $\mathcal{E}(\rho^{\otimes N})=N\mathcal{E}(\rho)$.

It is encouraging that nontrivial 1-scalability is builtin in a widely used quantum coherence measure such as the $\ell_1$-norm \cite{coh}. 
The binomial series (\ref{binomial}) is the simplest form of non-additivity for functions which are compatible with the tensor product structure \cite{parisio}, but the general solution (\ref{maclaurin}) allows for more intrincate functional forms of 1-S measures. It would be interesting to test other measures, either analytically or numerically, for 1-scalability, or, more generally, $q$-scalability. 
\begin{acknowledgments}
The authors thank B\'arbara Amaral and Nadja Bernardes for a discussion on the topics addressed in this manuscript.  This work received financial support from the Brazilian agencies Coordena\c{c}\~ao de Aperfei\c{c}oamento de Pessoal de N\'{\i}vel Superior (CAPES), Funda\c{c}\~ao de Amparo \`a Ci\^encia e Tecnologia do Estado de Pernambuco (FACEPE), and Conselho Nacional de Desenvolvimento Cient\'{\i}fico  e Tecnol\'ogico through its program CNPq INCT-IQ (Grant 465469/2014-0).
\end{acknowledgments}

\appendix
\begin{widetext}
	
\section{3rd order coefficient solution}\label{appendix:3order}

To solve the recurrence relation (\ref{3r}), we choose $K=a$ and apply the known solutions for $d_1(a^n)$ and $d_2(a^n)$ \cite{parisio}:
\begin{equation}
d_3(a^n)=a^{\nu(n-1)}d_3(a) + 2[d_2(a)]^2a^{\nu(n-1)}\left(\frac{a^{\nu(n-1)}-1}{a^\nu-1}\right) + d_3(a^{n-1})a^{3\nu}. \nonumber
\end{equation}
Remember that $a^n=N$. The recursive substitution of $d_3(a^{n})$ into itself $l=n-1$ times gives:
\begin{eqnarray}
d_3(N)&=&d_3(a)\left[a^{\nu (n-1)}+a^{\nu(n-2)}a^{3\nu}+\dots \right] + 2[d_2(a)]^{2} \sum_{l=1}^{n-1} \left\lbrace a^{\nu(n-l)} \left(\frac{a^{\nu(n-l)}-1}{a^\nu-1}\right) a^{3\nu(l-1)} \right\rbrace \nonumber \\ \nonumber \\ 
\dots&=&d_3(a)\sum_{l=1}^{n}a^{\nu(n-l)} a^{3\nu (l-1)} + \frac{2[d_2(a)]^2} {a^\nu-1}\left(\frac{N}{a^3}\right)^\nu\sum_ {l=1}^{n-1}a^{2\nu l}\left(N^{\nu}a^{-\nu l}-1\right).\nonumber \\ \nonumber
\end{eqnarray}
Redefining all sums to start at $l=0$ and using the geometric series $\sum_{l=0}^{n-1}x^l=\frac{1-x^{n}}{1-x}$ we get the 3th order in (\ref{maclaurin}):
\begin{equation}
d_3(N)=d_3(a)\left(\frac{N}{a}\right)^{\nu}\left(\frac{1-N^{2\nu}}{1-a^{2\nu}}\right) + 2[d_2(a)]^2 \left(\frac{N}{a^{2}}\right)^\nu \frac{(N^\nu-1)(N^\nu-a^\nu)}{(a^\nu-1)(a^{2\nu}-1)}\nonumber
\end{equation}

\section{Binomial coefficients}\label{bicoeff}
We start with (\ref{coef}) for $K=2$ (here the index $l$ is rewritten as $j-k$):
\begin{eqnarray}
d_j(N) &=& \sum_{k=0}^{j-1} d_{j-k}(N/2) \sum_{i=1}^{\begin{tiny}{{j-1} \choose {j-k-1}}\end{tiny}}\pi_i(j,j-k;2) \label{B1}
\end{eqnarray}
Remember that $N \in\mathds{P}_{2}$, so $\frac{N}{2}$ is an integer. The hypotesis that only $d_1(2)$ and $d_2(2)$ are non-zero means that $\pi_i(j,l;2)$ is a product of combinations of these two quantities only, so we need to consider the composition of $j$ using only the numbers $1$ and $2$, which has $\tiny{\frac{(j-k)!}{k!(j-2k)!}}$ elements, for $k$ repetitions of the number $2$ (the maximum value of $k$ is $\lfloor \frac{j}{2} \rfloor$), so:
\begin{eqnarray}
j \text{ } = \text{ } \underbrace{1+\dots+1}_{j-2k\text{ times}}+\underbrace{2+\dots+2}_{k\text{ times}} \text{ }\text{ }\text{ }\text{ }\text{ } \rightarrow \text{ }\text{ }\text{ }
\sum_{i=1}^{\begin{tiny}{{j-1} \choose {j-k-1}}\end{tiny}} \pi_i(j,j-k;2) = {{j-k} \choose {k}} [d_1(2)]^{j-2k} [d_2(2)]^{k}. \label{B2}
\end{eqnarray}
Putting (\ref{B2}) into (\ref{B1}) and considering the case $E^{(a=2)}(e)=2e+d_2(2)e^2$, we find the following recurrence relation:
\begin{eqnarray}
d_j(N) &=& \sum_{k=0}^{\lfloor \frac{j}{2} \rfloor} d_{j-k}(N/2){{j-k} \choose {k}} 2^{j-2k} [d_2(2)]^{k}. \nonumber
\end{eqnarray}
Now we test expression $d_j(N)=[d_2(2)]^{j-1}\tiny{{{N} \choose {j}}}$, induced in (\ref{induced}). Using the \textit{subset-of-a-subset property} \cite{combinatorial} we get:
\begin{eqnarray}
d_j(N) = [d_2(2)]^{j-1}\sum_{k=0}^{\lfloor \frac{j}{2} \rfloor} {{N/2} \choose {k}}{{N/2-k} \choose {j-2k}} 2^{j-2k}. \label{beforecauchy}
\end{eqnarray}
For the evaluation of the sum in (\ref{beforecauchy}) we take the integral representation $\tiny{\left( \begin{array}{c} n \\ m \end{array} \right)} = \frac{1}{2\pi i} \oint_{\Gamma} \frac{(1+z)^n}{z^{m+1}} dz$ to solve the problem using the Egorychev method \cite{egorychev,egorychevmethod}, where $z$ is a complex variable and $\Gamma$ is a small closed contour around $z=0$:
\begin{eqnarray}
\sum_{k=0}^{\lfloor \frac{j}{2} \rfloor} {N/2 \choose k}{{N/2-k} \choose {j-2k}} 2^{j-2k} 
= \sum_{k=0}^{\lfloor \frac{j}{2} \rfloor}\frac{1}{2 \pi i} \oint_{\Gamma} dz \frac{(1+z)^{N/2-k}}{z^{j-2k+1}}{N/2 \choose k}2^{j-2k}. \nonumber
\end{eqnarray}
Is easy to see that for $\lfloor \frac{j}{2} \rfloor<k<\frac{N}{2}$ the integral vanishes. By making $k=\lfloor \frac{j}{2} \rfloor+l$ (so $l \leq \frac{N}{2}- \lfloor \frac{j}{2} \rfloor$) the integrand does not have residue at $z=0$ for any value of $l \in (1,\frac{N}{2}- \lfloor \frac{j}{2} \rfloor)$ and, therefore, we can rewritte the sum with upper limit equal to $\frac{N}{2}$. Then we can use the binomial theorem:
\begin{eqnarray}
\frac{1}{2 \pi i} \oint_{\Gamma} dz \frac{(1+z)^{N/2}}{z^{j+1}}2^j \sum_{k=0}^{\frac{N}{2}}{N/2 \choose k} \left( \frac{z^2}{4(1+z)} \right)^k &=&
\oint_{\Gamma} dz \frac{{(1+z)^{N/2}}}{z^{j+1}}2^j \left[\frac{(z+2)^2}{4{(1+z)}}\right]^{\frac{N}{2}}. \nonumber
\nonumber
\end{eqnarray}
After some cancellations we make a simple change of variables $z\text{ }\rightarrow \text{ }2z'$ ($\Gamma \rightarrow \Gamma'$) and the integral becomes exactly the Cauchy integral representation of the binomial coefficient, so the proposition is proven. 
\begin{equation}
d_j(N) = [d_2(2)]^{j-1} {N \choose j}
\end{equation}
\end{widetext}


\begin{thebibliography}{99}
\bibitem{comm0} The term ``resource'' is employed in the general sense, since the
results to be derived do not rely on all the requirements for a quantity to be 
a resource (in the resource-theoretic sense).
\end{thebibliography}
\end{document}